\documentclass[aps,prb,reprint,superscriptaddress]{revtex4-2}
\usepackage{hyperref}
\usepackage{graphicx}
\usepackage{color}
\usepackage{amsfonts}
\usepackage{amsmath}
\usepackage[utf8]{inputenc}
\usepackage[T1]{fontenc}
\usepackage{mathtools}
\usepackage{bbm}
\usepackage{braket}
\usepackage{dsfont}
\usepackage[dvipsnames]{xcolor}
\hypersetup{colorlinks=true, urlcolor=blue, citecolor = blue}
\usepackage{amsmath} 
\usepackage{textcomp}
\usepackage{siunitx}
\begin{document}

\title{Emergent spin accumulation in non-Hermitian altermagnets}

\author{J. H. Correa}
\email{jorgehuamani90@gmail.com}
\affiliation{Universidad Tecnológica del Perú, Dirección de Investigación Región Sur}

\author{M. P. Nowak}        
\affiliation{AGH University of Krakow, Academic Centre for Materials and Nanotechnology, al. A. Mickiewicza 30, 30-059 Krakow, Poland.}

\author{A. Pezo}
\email{apezol@issp.u-tokyo.ac.jp}
\affiliation{Institute for Solid State Physics, University of Tokyo, Kashiwa, 277-8581, Japan}

\begin{abstract}

The recent interest in non-Hermitian (NH) systems has significantly broadened their application across condensed matter physics, offering a unique framework to explore out-of-equilibrium phenomena. Simultaneously, altermagnets have emerged as a distinct magnetic class, characterized by unconventional spin-split bands protected by crystal symmetries. In this work, we investigate the interplay between non-Hermitian dynamics and spin transport in these materials, focusing on the Edelstein effect. We demonstrate that the introduction of non-Hermiticity in $d$-wave altermagnets and $p$-wave unconventional magnets opens novel susceptibility components that are inaccessible in Hermitian counterparts. Our analysis reveals that these susceptibility channels are highly sensitive to the underlying symmetry of the order parameter. Crucially, our results show that the non-conservative nature of the system leads to the selective gain and loss of specific spin components, a phenomenon that can be tuned by the interplay between dissipation and the altermagnetic order. These components exhibit a distinct gain/loss profile that depends strictly on the N\'eel vector orientation, providing a new route for manipulating spin degrees of freedom through controlled non-conservative processes in emerging magnetic materials

\end{abstract}
\maketitle

\newcommand\red[1]{{\color{red}#1}}





\section{Introduction}

Altermagnets (AM), a new magnetic phase with coexistence of ferromagnetic and antiferromagnetic properties  with
spin polarization but zero net magnetization, has drawn attention as candidates for next-generation superconducting spintronic devices and promising applications in areas such as valleytronics, conductivity, and straintronics \cite{libor2022_prx, libor2022,Krempasky2024}. Within this framework, compounds such as CrSb, MnTe, and Mn$_5$Si$_3$ have been theoretically and experimentally identified as hosting remarkable altermagnetic signatures, including the strain-driven Anomalous Hall Effect (AHE) and the efficient generation of spin currents via spin-orbit torques \cite{sheoran2025, schwartz2025}, non-linear Edelstein effect \cite{jeron2025}, topological superconductivity \cite{ezawa2024, cayao2025}, Josepshon properties and diode effect \cite{lu2024, carlo2023, jacob2023,banerjee2024}. Beyond conventional AM, recent studies on $p$-wave unconventional magnets (UM) have revealed intriguing properties arising from their preserved time-reversal symmetry and broken inversion symmetry- features that clearly distinguishes them from AM materials \cite{birk2023, bhowal2025}. Notably, phenomena such as the emergence of the finite out-of plane Edelstein effect \cite{ezawa_pwave2025}, unique transport properties in hybrid systems \cite{morteza2025, linder2024_pwave, maeda} and unconventional superconductivity \cite{fukaya2025} have gained significant attention.

Despite this progress, the properties of AM(UM) are still largely restricted to their Hermitian features, effectively treating these systems as closed and dissipationless. However, devices made of these structures operate in open environments, where coupling to external leads, disorder, and finite quasi-particle lifetimes are unavoidable \cite{ashida2020}. These effects naturally introduce gain and loss processes, where a NH
description is necessary.

\indent The inclusion of NH effects has emerged as a promising avenue for exploring unconventional topological phases markedly different from their Hermitian counterparts \cite{gong2018}. These effects provide a realistic description of systems characterized by energy gain and loss, leading to complex energy spectra and new phenomena under boundary conditions \cite{emil2021}. In particular, eigenstates can exhibit strong localization, a phenomenon known as the NH skin effect  \cite{review_nonhermitian_skin, kazuki2019, kawabata2019, okuma2023}. Furthermore, NH systems feature exceptional points (EP), where eigenvalues and eigenstates coalesce—this is a fundamental aspect of NH physics \cite{emil2021}. Recently, EP were identified in $d$-wave altermagnets coupled to ferromagnetic leads, tuned by external magnetic fields \cite{dash2025, reja2024}. Quantum transport studies within the NH framework have further revealed significant dissipation effects in the Hall conductance \cite{nagaosa2023, chen2018}. Such new concepts have led to an increasing interest in understanding how NH properties can be used as new ways to manipulate the spin degree of freedom. Advances in material growth and experimental techniques have enabled detailed exploration and utilization of spin dynamics. This can be probed either by using spin pumping techniques \cite{Anadn2025}, exploiting the Seebeck effect \cite{7452553} or by considering a less invasive approach as it is the case of THz emission in spintronic devices \cite{THz_9508905,Rouzegar_2025,pezo2025spinorbitalrashbaresponse}. In particular, AM (UM), due to their anisotropic properties, offer an intriguing platform for the exploration of active spin–charge converters under various conditions. Their defining characteristic is that the conversion process does not depend on strong relativistic spin–orbit coupling \cite{chakraborty2025}, where collinear but alternating spin moments are responsible for mimicking the responses typically associated with heavy-element systems. This unique origin not only allows altermagnets to serve as efficient sources and detectors of spin currents but also embeds their functionalities in a manner dependent on crystal symmetry, leading to anisotropic and tunable effects \cite{32zc-ggjy}. 

\begin{figure}
    \centering
    \includegraphics[width=1.0\linewidth]{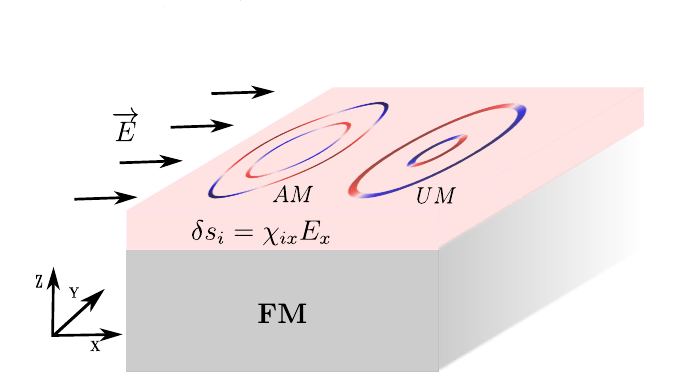}
    \caption{Schematic representation of the in-plane electric field effect at the AM(UM)/FM lead interface. The diagram illustrates the coupling between the external electric field $\vec{E}$ and the interface formed by a ferromagnetic lead (FM) and a unconventional magnet. The respective Fermi surfaces under Rashba SOC are shown, highlighting the $d$-wave symmetry (AM) and the $p$-wave symmetry (UM). The electric field in $\hat{x}$ direction gives the spin accumulation in $i$ direction accordingly with $\delta s_i = \chi_{ix} E_x$.
    \label{setup}}
\end{figure}

\noindent In this work, we investigate the spin Edelstein effect within a NH framework given by the coupling of an AM(UM)/FM lead interface as depicted in Fig. \ref{setup}. By employing a tight-binding (TB) formalism, we extend linear response theory to incorporate NH effects, identifying specific regions in phase space where the interplay between damping and gain yields experimentally accessible signatures. The structure of this article is as follows, In Sec. \ref{sec2}, we introduce the TB model for the NH altermagnetic system and derive the susceptibility tensors using the Kubo formalism and Green’s function techniques. Sec. \ref{sec3} present the result where we discuss the impact of N\'eel vector orientation and vertex corrections on our results. Finally, Sec. \ref{sec4} summarizes our findings and provides concluding remarks.

\label{sec1}

\section{Theoretical Model}
\label{sec2}

\subsection{Tight-binding Hamiltonian models}

Non-Hermiticity provides a natural framework for describing open quantum systems where dissipation and finite quasiparticle lifetimes—inherent to most experimental setups—play a fundamental role. Here, following Ref. \cite{cayao2023}, we introduce NH features by interfacing a AM(UM) system with a ferromagnetic lead, such coupling is given by the self-energy $\Sigma^{R}(\omega=0) = -i\Gamma\sigma_0  - i\gamma\sigma_z$ where $\Gamma = (\Gamma_{\uparrow}+ \Gamma_{\downarrow})/2$ and $\gamma = (\Gamma_{\uparrow}- \Gamma_{\downarrow})/2$, being
$\Gamma_{\sigma}=\pi(t^{'})^2\rho^{\sigma}_L$, $t^{'}$ is the value of the interfacial coupling in the ferromagnetic lead and $\rho^{\sigma}_L$ the spin dependent surface density of states, finally, $\sigma_0$ is the identity matrix and $\sigma_z$ the $z-$component Pauli matrix  \cite{cayao2023}

The resulting NH Hamiltonian is defined as $H_{NH} =H_{H} + \Sigma(\omega = 0)$, where the Hermitian part $H_{H} = H_{K} + H_{R} + H_{AL}$ includes kinetic energy, Rashba spin-orbit coupling (SOC), and AM(UM) contribution respectively. The kinetic and Rashba SOC Hamiltonians in reciprocal space are given by

\begin{equation}
\begin{aligned}
H_{K}(\mathbf{k}) &= -2t\left(\cos k_x + \cos k_y\right)\sigma_0 - \mu \, \sigma_0, \\
H_{R}(\mathbf{k}) &= \lambda_R \left(\sin k_y \, \sigma_x - \sin k_x \, \sigma_y\right)
\end{aligned}
\label{hamiltonianos}
\end{equation}
where $t$ is the hopping parameter and $\mu$ is the chemical potential, likewise $\lambda_{R}$ represent the Rashba SOC parameter. For the altermagnetic sector, we adopt the following $d$-wave and $p$-wave Hamiltonians

\begin{equation}
\begin{aligned}
H^{d}_{AL}(\textbf{k}) &= \alpha [2\sin k_x \sin k_y \sin 2\theta \\
&+ (\cos k_x - \cos k_y) \cos 2\theta] (\hat{n} \cdot \vec{\sigma}) \\
H^{p}_{UM}(\textbf{k}) &= [t_x \sin k_x + t_y \sin k_y] (\hat{n} \cdot \vec{\sigma})
\end{aligned}
\label{hamiltonian_altermagnet}
\end{equation}

$H^d_{AL}$ describes a $d$-wave altermagnet with strength $\alpha$, where the orientation angle $\theta$ interpolates between $d_{x^2-y^2}$ ($\theta = 0$) and $d_{xy}$ ($\theta = \pi/4$) symmetries \cite{libor2022,rao2024}. Similarly, $H^p_{UM}$ represents a $p$-wave unconventional magnet \cite{birk2023, morteza2025, fukaya2025} with strengths $t_x$ and $t_y$ along the $p_x$ and $p_y$ symmetries, respectively. In both models, $\hat{n}$ denotes the orientation of the N\'eel vector \cite{ezawa_pwave2025}. Fig. \ref{setup}, shows the setup used in the model, where the Fermi surfaces for $d$-wave AM and $p$-wave UM without NH parameters show the spin-splitting accordingly to the value of the Rashba SOC and $\hat{n} = (0,0,1)$.

\subsection{Kubo formula extension for spin dynamics}
\label{kubo}
To investigate the out-of-equilibrium spin accumulation properties, we employ linear response theory. Within this framework, the electrical driven spin accumulation arises as the response to an applied in-plane electric field $E_{j}$ defined by the relation $\braket{\delta \hat{s}_{i}} = \sum_{j}\chi_{ij}E_{j}$. Using the Kubo formalism, the proportionality constant $\chi_{ij}$, also known as the spin Edelstein susceptibility tensor, can be written as \cite{manchon2020, ovalle2023}, 

\begin{equation*}
\chi_{ij}  = -\frac{e\hbar}{8\pi} \int \partial_{\epsilon} f(\epsilon)\,d\epsilon\hspace{0.1cm} \mathrm{Re}[\mathrm{Tr}\{ \hat{v}_i(G^{R-A}(\mathbf{k},\varepsilon))
\end{equation*}
\begin{equation}
\hat{\mathcal{\sigma}}_{j}(G^{R-A}(\mathbf{k},\varepsilon))\}].
\label{extrinsic}    
\end{equation}

Here, $G^{R \pm A} = G^{R} \pm G^{A}$, where $G^{R}$ and $G^{A}$ denote the retarded and advanced Green's functions (GFs), respectively. Similarly, $\hat{v}_i = \partial_{k_{i}} H_{NH}(\textbf{k})$ and $\hat{\sigma}_{j}$ represent the momentum-dependent velocity and spin matrices. The term $f(\epsilon)$ denotes the Fermi-Dirac distribution, and $\mathrm{Tr}$ indicates the trace over the internal degrees of freedom. Moreover, we restrict our analysis to the low-temperature limit ($T \rightarrow 0$), where the derivative of the distribution function reduces to a delta function, $\partial_{\epsilon} f(\epsilon) \rightarrow \delta(\epsilon - \mu)$. In terms of response's symmetry, upon the action of Time-Reversal symmetry ($\mathcal{T}$), Eq. \ref{extrinsic} is $\mathcal{T}-$even. 

To incorporate non-Hermiticity arising from the coupling to the ferromagnetic lead, we adopt the biorthogonal basis formalism \cite{nagaosa2023, chen2018}. The retarded NH Green's function is expressed as 

\begin{equation}
    G^{R}(\mathbf{k},\varepsilon) = \frac{P_{+}(\mathbf{k})}{\varepsilon - E^{R}_{+}(\mathbf{k})} + \frac{P_{-}(\mathbf{k})}{\varepsilon - E^{R}_{-}(\mathbf{k})}
\label{green_retarded}
\end{equation}
Here, $P_{\pm}$ are the projection operators onto the eigenstates, defined as $P_{\pm} = \ket{\Psi_{\pm}^{R}}\bra{\Psi_{\pm}^{L}}$, where $\ket{\Psi_{\pm}^{R,L}}$ denote the right(left) eigenvectors obtained from our NH Hamiltonian and satisfied the biorthogonal condition  $\bra{\Psi_m}\ket{\Psi_{n}} = \alpha_{nm}$ where ensures $\sum_{\pm} P_{\pm} = \sigma_0$ \cite{chen2018}, likewise,  $E_{\pm}$ are the corresponding complex energies obtained from the NH Hamiltonain. The integrand is performed over the BZ and we only take the real part of the susceptibility tensor. From Eq.~\ref{green_retarded} the advanced GF is given by $G^{A} = [G^{R}]^{\dagger}$. In the Hermitian limit $\Sigma^{R}(\omega = 0) \rightarrow 0$, the GF recovers their Hermitian form with $\ket{\Psi_{\pm}^{R}} = \ket{\Psi_{\pm}^{L}}$.

\section{Results}
\label{sec3}

\noindent In this section, we investigate the field-driven spin accumulation considering an external electric field applied along the $x$ direction i.e $\mathbf{E} = E_x \hat{x}$ to the system defined by the AM(UM)/ FM lead coupling (see Fig. \ref{setup}). Specifically we calculate the spin accumulation along the  $j$ direction, given by Eq.\ref{extrinsic}. The real part of the susceptibility tensor is expressed in units of $\chi_{0} = e\hbar/\pi$ and unless otherwise specified, all numerical calculations are performed with the parameters $t=0.5$, $\lambda_{R}=0.3$ and $\mu = 0$.

\subsection{Out-of-plane Néel order $\hat{n} \parallel z$ direction}
\begin{figure}
    \centering
    \includegraphics[width=1.0\linewidth]{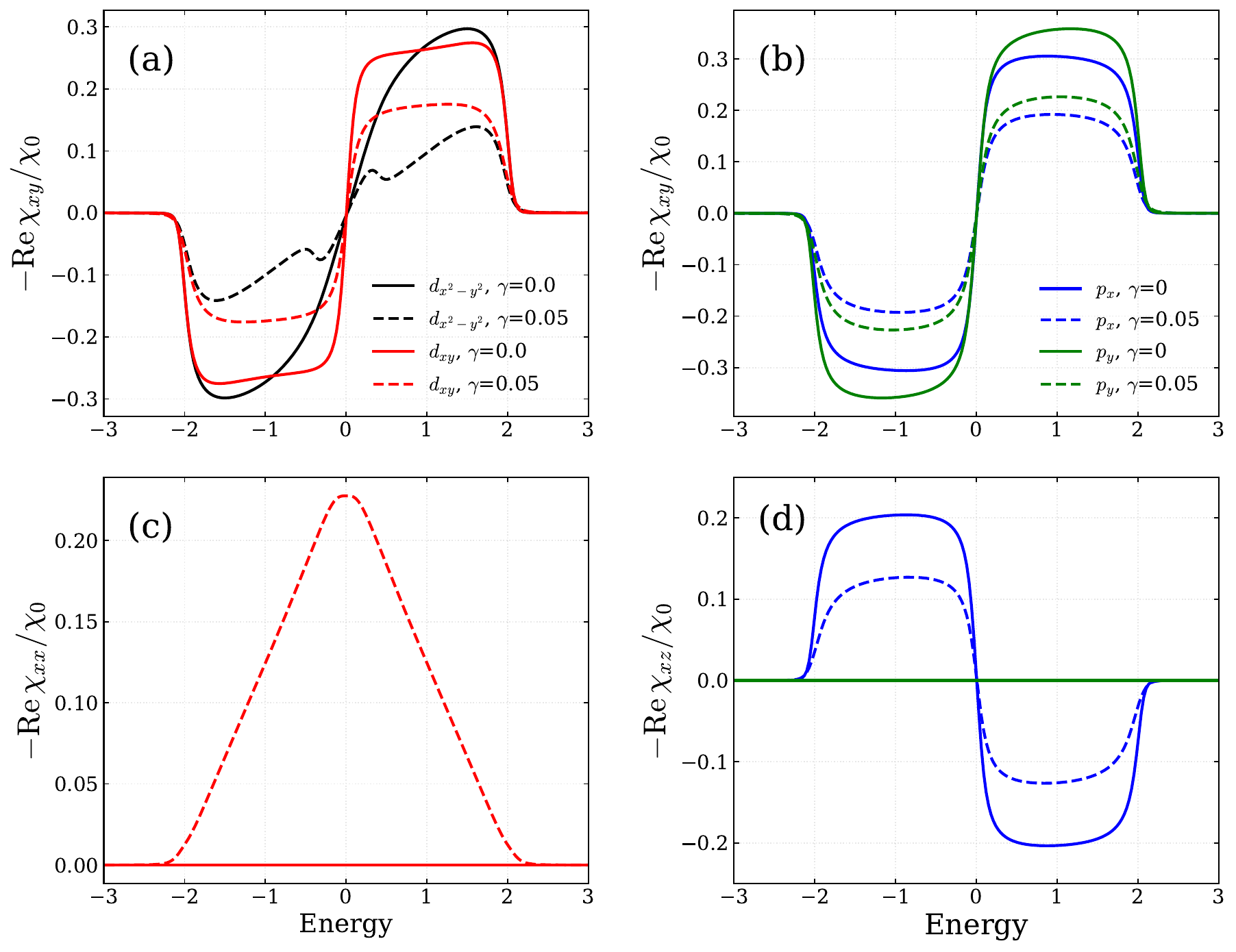}
    \caption{Real part of the spin susceptibility tensor for a NH systems with Néel vector along the $z$ direction. Left panel [(a), (c)]: $ d$-wave AM and Right panel [(b), (d)] : $p$-wave UM. The transversal component $\chi_{xy}$ [(a) and (b)] decreases for $\gamma \neq 0$ while the diagonal component $\chi_{xx}$ arise for $d_{xy}$ wave (c) and an out-of plane component $\chi_{xz}$ decreases by the action of $\gamma$ (d).  For (a) and (c), $\alpha = 0.2$, while for (b) and (d), $t_{x} = 0.2, t_y=0 $ and $t_x=0, t_y =0.2$ for $p_x$ and $p_y$ respectively.}
    \label{conductance_dwave_pwave_nonhermitian_neelz}
\end{figure}

In Fig.~\ref{conductance_dwave_pwave_nonhermitian_neelz} we present the corresponding components of the spin Edelstein susceptibility responses for $d$-wave [(a), (c)] and $p$-wave [(b), (d)] symmetries in left and right panels respectively considering a N\'eel vector oriented along the $z$ direction i.e. $\vec{n} = (0 , 0, 1) $. The Hermitian case is represented by solid curves ($\gamma = 0$), while the NH regime is represented by dashed curves with $\Gamma = \gamma = 0.05$, and this value aligns with previous studies \cite{cayao2023, elsa2025}.

Focusing first on the $d$-wave AM, the transversal in-plane susceptibility $\chi_{xy}$ exhibits an almost oscillatory behavior that is symmetric around $E = 0$. While this response—enhanced by Rashba SOC—is consistent with recent reports~\cite{jeron2025,chakraborty2025}, the symmetry $C_{4z}\mathcal{T}$ ensuring that both, the longitudinal ($\chi_{xx}$) and out-of plane ($\chi_{xz}$) components vanish. Revisiting it here is instructive, as its physical origin becomes especially transparent in the  NH regime.

The inclusion of non-Hermiticity, controlled by the parameters $\gamma$ and $\Gamma$, modifies the intensity of $\chi_{xy}$, as evidenced by the dashed curves. For the $d$-wave case, the magnitude of the susceptibility is progressively reduced as $\gamma$ increases. This suppression signals dissipative effects characteristic of NH quantum transport ~\cite{michen2022,nagaosa2023}. Remarkably, non-Hermiticity qualitatively alters the symmetry-imposed absence of longitudinal components. As shown in Fig.~\ref{conductance_dwave_pwave_nonhermitian_neelz}(c), a finite longitudinal in-plane response $\chi_{xx}$ emerges for the $d_{xy}$ altermagnet (red dashed curve) upon activating $\gamma$. Unlike the dissipative suppression observed in $\chi_{xx}$, the $\chi_{xx}$ signal is significantly enhanced with increasing $\gamma$. This suggests a NH gain-loss mechanism where the complex self-energy redistributes the spin-dependent spectral weight, effectively amplifying one spin channel while another undergoes dissipation. The emergence of this component can be understood through the momentum parity of the integrand within the BZ as will be shown in Sec. \ref{effective_model}. In the $d_{xy}$ case, the product of the momentum-dependent velocities and the altermagnetic order parameter—proportional to $\sin k_{x}\sin k_y$ —acquires an overall even parity only in the presence of the non-Hermitian coupling $\gamma$, consequently, the integration yields a non-vanishing contribution. In contrast, for the $d_{x^{2}-y^{2}}$ 
symmetry, the mismatch between the velocity vectors and the $d$-wave AM form factor preserves the null response, even in the NH regime. On the other hand, the out-of-plane component $\chi_{xz}$ is negligible compared to $\chi_{xx}$ that appears three orders of magnitude lower and only with large values of NH parameters (not shown here). 

We now turn our attention to the $p$-wave symmetry. Similar to the $d$-wave, the system exhibits finite $\chi_{xy}$ which decreases with the action of $\gamma$ [see Fig. \ref{conductance_dwave_pwave_nonhermitian_neelz} (b)], but in contrast, the $p_x$ UM shows a finite $\chi_{xz}$ even in the Hermitian regime, as shown in Fig.~\ref{conductance_dwave_pwave_nonhermitian_neelz}(d) by the blue solid curve which agree with Ref \cite{ezawa_pwave2025}. Upon increasing $\gamma$, this response is progressively suppressed, as indicated by the blue dashed curve, displaying a dissipative behavior similar to that observed for $\chi_{xy}$. Symmetry considerations dictate that this response is highly sensitive to the $p$-wave direction; for an electric field applied along the $x$-direction, the response is finite for the $p_x$
symmetry but vanishes for $p_y$. Conversely, an electric field in the $y$-direction would activate the $p_y$ channel while rendering the $p_x$ response null. Notably, while activating $\gamma$ induces a spin population in the $d$-wave regime, it is insufficient to lift the parity in the $p$-wave $\chi_{xx}$ 
component.

\subsection{In-plane Néel configurations $\hat{n} \parallel x, y$}

\begin{figure}
    \centering
    \includegraphics[width=1.\linewidth]{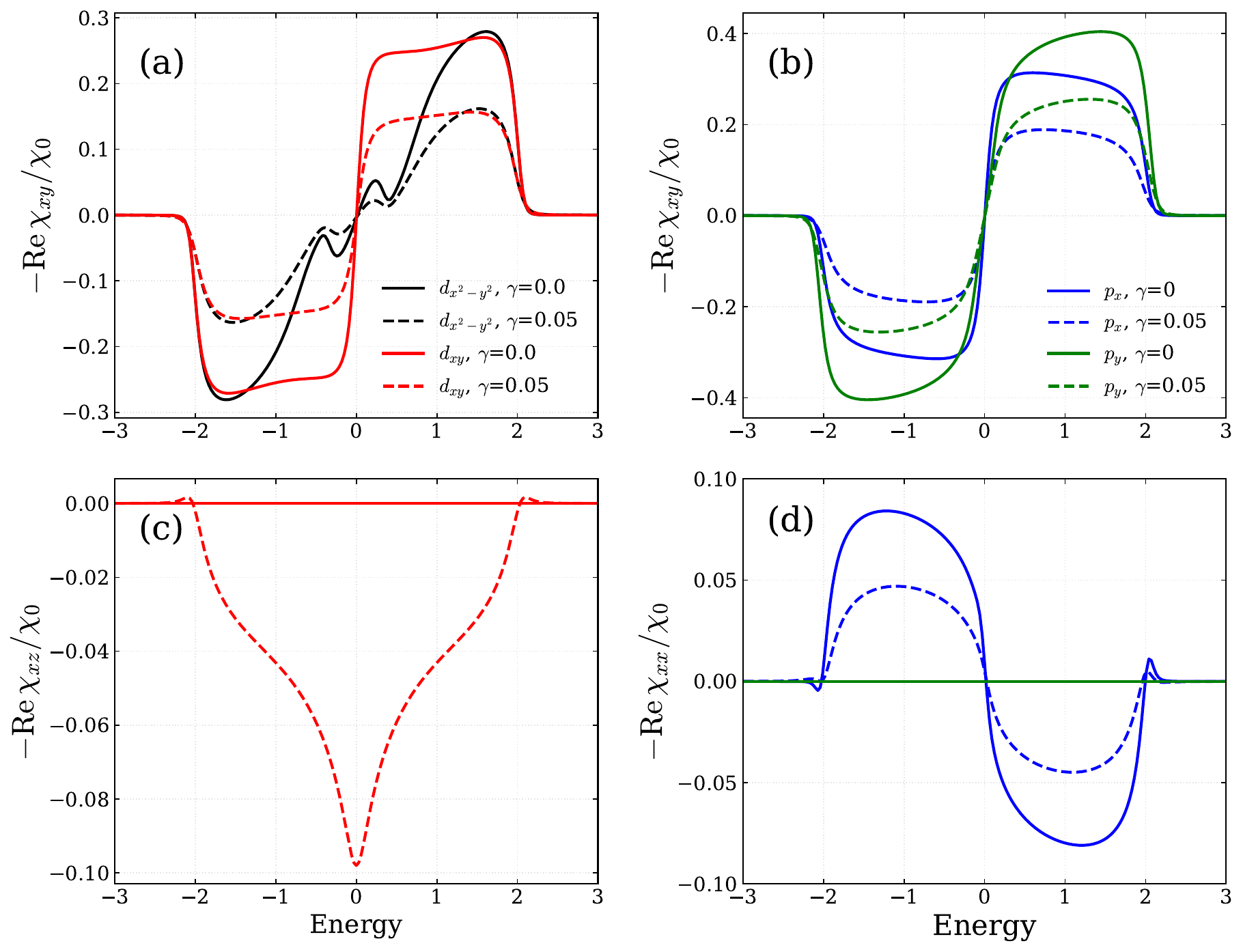}
    \caption{Real part of the spin susceptibility components for a non-Hermitian systems with Neel vector along $x$ direction. Left panel: $ d$-wave AM [(a), (c)] and Right panel : $p$- wave UM [(b), (d)]. The transversal component $\chi_{xy}$ [(a) and (b)] decreases for $\gamma \neq 0$, finite out-of plane component $\chi_{xz}$ arise for $d_{xy}$ wave (c) and longitudinal component $\chi_{xx}$ decreases by the action of $\gamma$ (d).  For (a) and (c), $\alpha = 0.2$, while for (b) and (d), $t_{x} = 0.2, t_y=0 $ and $t_x=0, t_y =0.2$ for $p_x$ and $p_y$ respectively.}
    \label{conductance_dwave_pwave_nonhermitian_neelx}
\end{figure}

After discussing the classic N\'eel vector orientation, we move to the N\'eel vector oriented along the $x$ direction $\vec{n} = (1, 0, 0)$. In Fig.~\ref{conductance_dwave_pwave_nonhermitian_neelx} we plot the susceptibility components for $d$-wave AM [(a), (c)] and $p$-wave UM [(b),(d)] in left and right panels, respectively. Similar to the previous case, the transversal in-plane component $\chi_{xy}$ exhibits an almost oscillatory behavior in both systems, with its magnitude decreasing as $\gamma$ increases, as shown in panels (a) and (b).

Focusing on the $d$-wave AM and in the Hermitian limit, the components $\chi_{xx}$ and $\chi_{xz}$ vanish, where the rotation symmetry is still preserved even for different N\'eel vector. In contrast, in this case, the component $\chi_{xz}$ emerges in the NH regime only on the $d_{xy}$-wave AM, as illustrated by the dashed red curve in panel (c). This can be understood as the NH term produce a spin torque where the altermagnet parameter competes directly with the Rashba SOC parameter and the term coupled with $\sigma_z$ is purely imaginary $\sim i\gamma$, this term gives an even parity in the integrand and therefore the appareance of $\chi_{xz}$.

Regarding the $p$-wave symmetry, the $\chi_{xy}$ signal decreases with increasing $\gamma$, mirroring the dissipative trend observed in the $d$-wave case. However, a key distinction arises in the $\chi_{xz}$ component only for $p_x$-wave AM, which is already finite in the Hermitian regime [Fig.~\ref{conductance_dwave_pwave_nonhermitian_neelx}(d), blue solid curve], consistent with Ref.~\cite{ezawa_pwave2025}. In this sector, the NH term acts primarily as a dissipator, progressively suppressing the signal intensity (blue dashed curve). This contrast with the $d$-wave AM case—where NH terms act as a generative mechanism for new components—can be understood through the parity of the orbital form factors. The $p_x$ symmetry possesses odd parity in only one momentum direction; consequently, the overall parity of the integrand is determined by the remaining velocity and spin operators, which allows for a non-vanishing integral even without NH framework. In contrast, the $d_{xy}$
symmetry ($\propto \sin k_x \sin k_y$) is odd in both directions, enforcing a global cancellation in the Hermitian limit that can only be lifted by the complex spectral redistribution induced by $\gamma$.

\begin{figure}
    \centering
    \includegraphics[width=1.\linewidth]{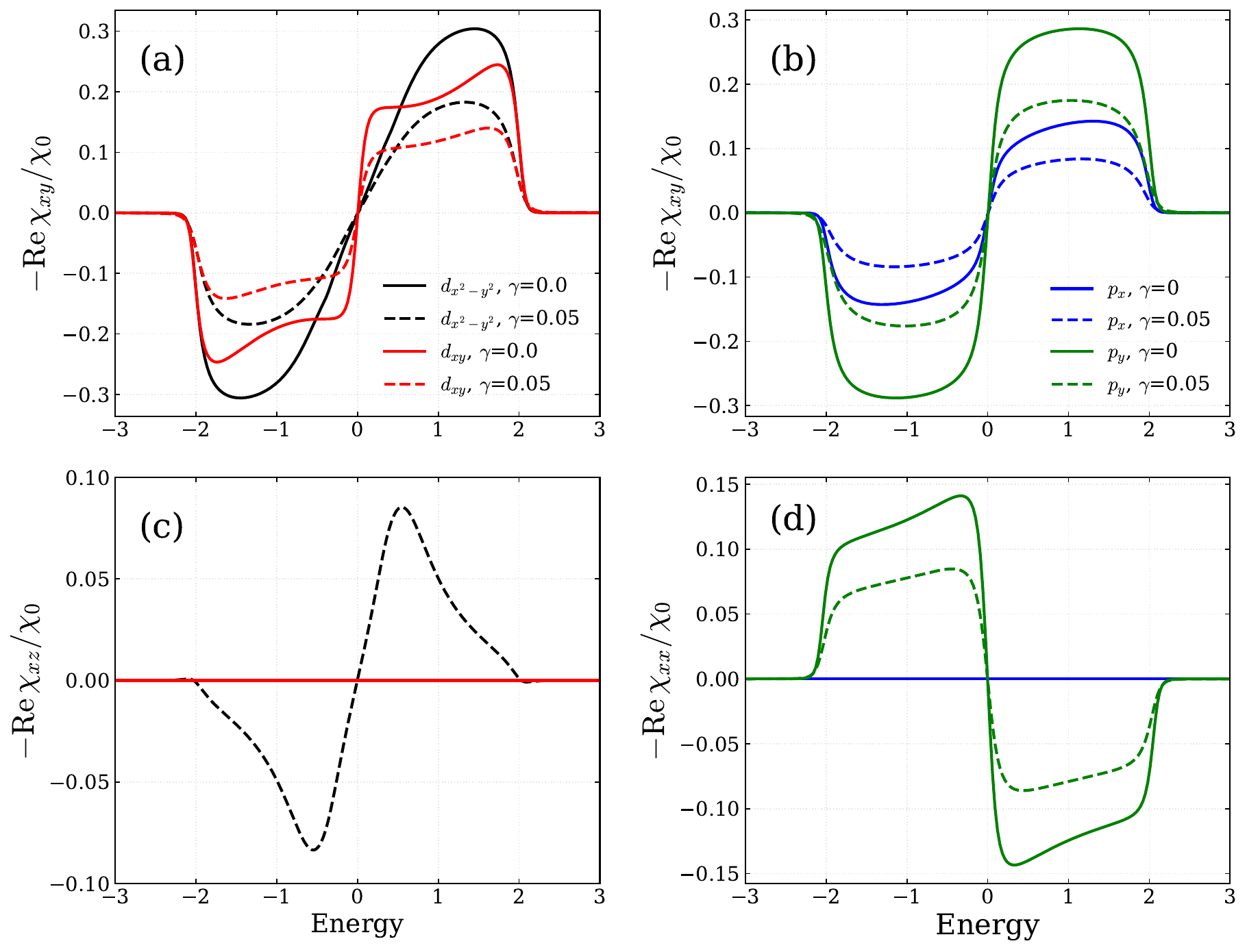}
    \caption{Real part of the spin susceptibility components for a non-Hermitian systems with Neel vector along $y$ direction. Left panel [(a), (c)]: $ d$-wave AM and Right panel [(b), (d)] : $p$-wave UM. The transversal component $\chi_{xy}$ [(a) and (b)] decreases for $\gamma \neq 0$, finite out-of plane component $\chi_{xz}$ arise for $d_{x^{2}-y^{2}}$ wave (c) and longitudinal component $\chi_{xx}$ for $p_y$ decrease by the action of $\gamma$ (d).  For (a) and (c), $\alpha = 0.2$, while for (b) and (d), $t_{x} = 0.2, t_y=0 $ and $t_x=0, t_y =0.2$ for $p_x$ and $p_y$ respectively.}
    \label{conductance_dwave_pwave_nonhermitian_neely}
\end{figure}

On the other hand, the situation undergoes a significant change when the N\'eel vector is oriented along the $y$-direction, i.e., $\hat{n}=(0, 1, 0)$, as illustrated in Fig.~\ref{conductance_dwave_pwave_nonhermitian_neely}. Similar to the previous case, the $\chi_{xy}$ component in both $d$-wave AM and $p$-wave UM systems [(a) and (b)] exhibits a symmetric, oscillatory-like profile whose magnitude is progressively suppressed by $\gamma$. This confirms that the $\chi_{xy}$ response remains robust and finite regardless of the N\'eel vector orientation. However, the NH coupling $\gamma$ induces a qualitative change in the other susceptibility components. Specifically, for $\hat{\mathbf{n}} \parallel y$, the $d_{x^2-y^2}$ altermagnet exhibits the emergence of a finite $\chi_{xz}$ response in the NH regime [Fig.~\ref{conductance_dwave_pwave_nonhermitian_neely}(c)]. Remarkably, $\chi_{xz}$ remains vanishingly small for this symmetry, indicating that the $y$-axis orientation shifts the NH-induced activation exclusively to the collinear sector for the $d_{x^2-y^2}$ symmetry. On the other hand, the $y$-orientation leads to a finite $\chi_{xx}$ in the $p_y$-wave UM even in the Hermitian limit, as indicated by the solid green curve in (d); this result is in good agreement with Ref.~\cite{ezawa_pwave2025}. The intensity of this component decreases with increasing $\gamma$, as shown by the green dashed curve.

\subsection{Analysis of effective Hamiltonian}
\label{effective_model}
In this section, we derive an analytical expressions for the trace of the response susceptibility matrix defined in Eq. \ref{extrinsic}. In order to simplify the calculations we will employ the effective models of Eqs. \ref{hamiltonianos} and \ref{hamiltonian_altermagnet} within a $\mathbf{k} \cdot \mathbf{p}$ approximation \cite{jeron2025, libor2022}. Furthermore, we examine the resulting momentum-space spin textures, specifically characterizing their behavior within the NH regime to assess the impact of dissipation on the spin accumulation. Throughout this analysis, we focus on the configuration where the N\'eel vector is aligned along the $z$ axis; however, the formal framework presented here is general, and extending the analysis to other N\'eel vector orientations is straightforward.

\subsubsection{Spin accumulation for $d$-wave AM and out-of-plane Néel order $\hat{n} \parallel z$ direction}

In order to demonstrate the emergence of 
$\chi_{xx}$ for the $d_{xy}$-wave AM, we computed the effective retarded GF of two level system \cite{hirsbrunner2019}, which is given by
\begin{align}
    G^{R}_{kp}(\mathbf{k},E) = [(E +i\eta)\mathbb{I} - H_{kp}]^{-1},
\end{align}
where $\eta$ is the broadening and the two-level $\mathbf{k} \cdot \mathbf{p}$ Hamiltonian can be written as $H_{kp} = d_0\sigma_0 + \mathbf{d}\cdot \mathbf{\sigma}$ where 
\begin{equation}
\begin{split}
    d_x &= \lambda_R k_y, \\
    d_y &= -\lambda_R k_x, \\
    d^{\theta=\pi/4}_z &= 2\alpha k_x k_y - i\gamma
\end{split}
\label{altermagnet_dxy}
\end{equation}
Based on \cite{kunst2019}, we can find an analytical expression for $G^{R}$ which can be written as 
\begin{equation}
    G^{R}_{kp}(\mathbf{k},\varepsilon) = \frac{\varepsilon \mathbb{I} + \mathbf{d}.\mathbf{\sigma}}{\varepsilon^{2} - d^{2}},
\label{green_effective}
\end{equation}
\newline
where $\varepsilon = E + i\eta$ and $d = \sqrt{d_x^2 + d_y^2 + d_z^2}$. After some algebraic steps, the effective retarded GF can be expressed as 
\begin{widetext}
\begin{align}
  G^{R}_{kp}(\textbf{k},E)= \frac{1}{D(\mathbf{k})} \begin{bmatrix}
-2\alpha k_x k_y - i(\eta -\gamma) - E & -\lambda(ik_x + k_y) \\
\lambda(ik_x - k_y) & 2\alpha k_x k_y - i(\eta + \gamma) - E
\end{bmatrix},
\end{align}
\end{widetext}
with $D(\mathbf{k}) = \lambda^{2}(k_x^2 + k_y^2) - (i\eta + E)^2 + (2\alpha k_x k_y - i\gamma)^2$. With this in hand, we can compute an analytical expression for the trace of the matrix in Eq. \ref{extrinsic}. For spin accumulation in the $x$ direction for momentum up to the third order and the NH parameter to the second order, we obtain.

\begin{widetext}
\begin{align}
\mathrm{Tr}^{d_{xy}}\!\left\{ \hat{v}_x G^{R-A}(\mathbf{k},\varepsilon)
\hat{\mathcal{\sigma}}_{x} G^{R-A}(\mathbf{k},\varepsilon) \right\}_{kp}
&= \frac{\gamma^2 \left(-128\alpha^2 k_x^3 k_y^3 \lambda^3 - 128\alpha^2 k_x k_y^3 \lambda E^2\right) + \gamma\left(-192\alpha \eta k_x^2 k_y^2 \lambda^3 E - 64\alpha \eta k_y^2 \lambda E^3\right)}{-16\alpha^2 k_x^2 k_y^2 E^6 + \gamma^{2}B + \gamma P + 12k_x^2 k_y^2 \lambda^4 E^4 - 4k_x^2 \lambda^2 E^6 - 4k_y^2 \lambda^2 E^6 + E^8},
\label{trace_kp_dxy}
\end{align}
\end{widetext}
\begin{widetext}
    \begin{align}
    B &= -16\alpha^2 k_x^2 k_y^2 E^4 + 24k_x^2 k_y^2 \lambda^4 E^2 - 12k_x^2 \lambda^2 E^4 - 12k_y^2 \lambda^2 E^4 + 4E^6,\\ 
    P &= -256\alpha^3 \eta k_x^3 k_y^3 E^3 + 64\alpha \eta k_x^3 k_y^3 \lambda^4 E - 64\alpha \eta k_x^3 k_y \lambda^2 E^3 - 64\alpha \eta k_x k_y^3 \lambda^2 E^3 + 32\alpha \eta k_x k_y E^5 \nonumber.
\end{align}
\end{widetext}

This analytical expression reveals that the spin accumulation in the $x$-direction vanishes when $\gamma=0$, a result consistent with our numerical TB calculations. Although the expansion is truncated at a specific order, the outcome in the Hermitian limit remains unchanged due to the symmetry of the momentum terms. In this limit, the trace is an odd function of the momentum  $k$, ensuring that the integral of $\chi_{xx}$ over the entire BZ is identically zero. However, for $\gamma \neq 0$, the non-Hermitian terms break this symmetry, shifting the parity of the function to even and leading to the emergence of a finite $\chi_{xx}$. A similar computation can be made by the symmetry of $d_{x^{2}-y^{2}}$ where in this case $d^{\theta=0}_{z} = \alpha (k_x^{2} -k_{y}^{2}) - i\gamma$ in Eq. \ref{altermagnet_dxy} and the trace is given by 

\begin{widetext}
\begin{align}
\mathrm{Tr}^{d_{x_{2}-y_{2}}}\!\left\{ \hat{v}_x G^{R-A}(\mathbf{k},\varepsilon)
\hat{\mathcal{\sigma}}_{x} G^{R-A}(\mathbf{k},\varepsilon) \right\}_{kp} = \frac{\gamma^2 G + \gamma H}{8\alpha^2 k_x^2 k_y^2 E^6 + \gamma^2 E + \gamma F + 12k_x^2 k_y^2 \lambda^4 E^4 - 4k_x^2 \lambda^2 E^6 - 4k_y^2 \lambda^2 E^6 + E^8},
\end{align}
\end{widetext}

\begin{widetext}
    \begin{align}
     G&= 128\alpha^2 k_x^3 k_y^3 \lambda^3 - 64\alpha^2 k_x^3 k_y \lambda E^2 + 64\alpha^2 k_x k_y^3 \lambda E^2\\ \nonumber
    H& = 384\alpha^3 \eta k_x^3 k_y^3 \lambda E - 64\alpha \eta k_x^3 k_y \lambda^3 E + 192\alpha \eta k_x k_y^3 \lambda^3 E - 64\alpha \eta k_x k_y \lambda E^3\\ \nonumber
    \beta&= 8\alpha^2 k_x^2 k_y^2 E^4 + 24k_x^2 k_y^2 \lambda^4 E^2 - 12k_x^2 \lambda^2 E^4 - 12k_y^2 \lambda^2 E^4 + 4E^6\\ \nonumber
    F& = 16\alpha \eta k_x^2 E^5 - 16\alpha \eta k_y^2 E^5 \nonumber.
\end{align}
\end{widetext}
From this expression, it is straightforward to verify that the parity of the numerator is odd, while the denominator is even, consequently, the entire trace behaves as an odd function of the momentum, ensuring that its integral over the BZ vanishes due to symmetry. As demonstrated by our analytical results, we confirm the emergence of finite $\chi_{xx}$ in $d_{xy}$ symmetry and remains zero in $d_{x^{2}-y^{2}}$ symmetry. Furthermore, a key distinguishing feature is the requirement of a finite altermagnetic parameter $\alpha$ to obtain a non-vanishing trace in Eq.~\eqref{trace_kp_dxy}, which clearly differentiates this system from a conventional Rashba semiconductor.

\subsubsection{Spin expectation values for $\hat{n}\parallel z $}
Although the biorthogonal approach could be used for these calculations, we will restrict our analysis to the right eigenstates as an approximation for small values of the NH parameter. The spin expectation values are defined
\begin{equation}
   S_i = \bra{\psi_{\pm}}\sigma_i\ket{\psi_{\pm}},
\end{equation}
where $\ket{\psi_{\pm}}$ are the eigenstates of the $\mathbf{k} \cdot \mathbf{p}$ Hamiltonian. With this in hand, the expectation values components can be written as 
\begin{equation}
\begin{aligned}
S_x = 2|N|^{2}\text{Re}[f_{k}], & \quad  S_y = 2|N|^{2}\text{Im}[f_{k}], \\
S_z &= |N|^{2}(1 - |f_{k}|^{2}),
\end{aligned}
\label{spin_texture}
\end{equation}
where 
\begin{equation}
|N|^{2} = \frac{1}{1 + |f_{k}|^{2}}, \hspace{0.2cm}f_{k} =  \frac{d_{z} + d}{-(d_x - id_y)}. 
\end{equation}

\begin{figure*}
    \centering
    \includegraphics[width=1.0\linewidth]{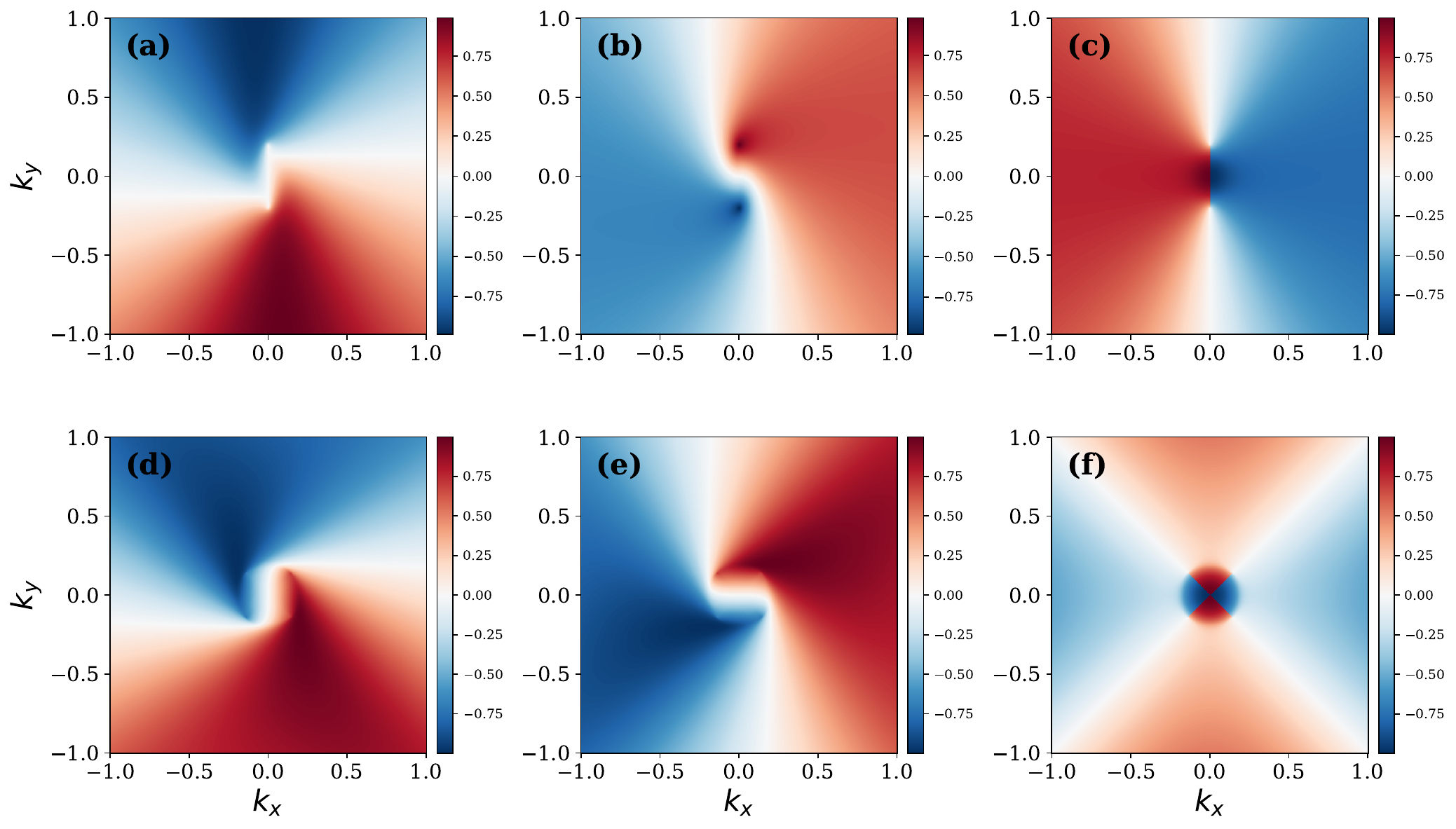}
    \caption{Momentum-space spin textures for $p_x$-wave and $d_{xy}$-wave magnetic models.
    The top panel displays the spin textures for the $p_x$-wave UM, while the bottom panel shows the results for the $d_{xy}$-wave AM. 
    From left to right, the columns represent the expectation values of the spin operators $S_{x}$ [(a), (c)], $S_{y}$  [(b), (d)], and $S_z$ [(c), (f)], respectively. In both cases the white line described where the texture is ill-defined. For $p_x$-wave UM, two EP is well pronunced in (b), while for $d_{xy}$- wave AM four EPs arise as can observed in (d) and (e).}
    \label{texture_spin_dwave_pwave_neelz}
\end{figure*}

The spin expectation values $S_i$ for the $p_x$-wave UM and $d_{xy}$-wave AM models are displayed in Fig. \ref{texture_spin_dwave_pwave_neelz} in top and bottom panels respectively. In panel (a), the longitudinal component exhibits a clear even parity with respect to the momentum $k_x$, such that $S_x(k_x, k_y) = S_x(-k_x, k_y)$. Under the application of an external electric field $\mathbf{E} = E_x \hat{x}$, this symmetry prevents a net spin accumulation; the contributions from opposite momenta cancel exactly, precluding the emergence of a longitudinal susceptibility $\chi_{xx}$.

In contrast, the transverse component $S_y$ in panel (b) obeys an odd parity relation, $S_y(k_x, k_y) = -S_y(-k_x, k_y)$. This antisymmetry allows the electric field to induce a population imbalance, thereby generating a finite $\chi_{xy}$. Notably, the NH coupling introduces EPs where the complex energy eigenvalues and their corresponding eigenstates coalesce, manifested as well-defined nodal structures in the texture. The positions of the EPs are determined by the diagonalization of $H_{kp}$. Following a procedure similar of the Ref.\cite{reja2024}, we find that EPs for the $p_{x}$-UM are located at:
\begin{equation}
    (k_{x}, k_{y}) = \left(0, \pm\frac{\gamma}{\lambda_R}\right).
\end{equation}
These points are clearly visible in Fig. \ref{texture_spin_dwave_pwave_neelz}(b) as the singularities where the spin texture orientation becomes ill-defined. As we move toward the boundaries of the Brillouin zone, the spin polarization decays rapidly, leading to the overall suppression of the integrated susceptibility $\chi_{xy}$ observed in Fig. \ref{conductance_dwave_pwave_nonhermitian_neelz}. A similar symmetry argument applies to the out-of-plane component $S_z$ in panel (c), which also satisfies $S_z(k_x, k_y) = -S_z(-k_x, k_y)$. However, the inversion of the spin-momentum locking compared to $S_y$ results in an opposite sign for the $\chi_{xz}$ response, consistent with the transport coefficients presented in Fig. \ref{conductance_dwave_pwave_nonhermitian_neelz}(d). 

On the other hand, the scenario changes significantly for the $d_{xy}$-AM model. In the Hermitian limit ($\gamma = 0$), the $\chi_{xx}$ component vanishes identically due to symmetry constraints. However, for $\gamma \neq 0$, four EPs emerge as reported in Refs.~\cite{reja2024, dash2025}, inducing a pronounced distortion of the spin textures in their vicinity. Unlike the case in panel (a), the $S_x$ component no longer exhibits perfect even parity with respect to $k_x$. Consequently, the application of an electric field $\mathbf{E} = E\hat{x}$ generates a spin population imbalance, leading to the emergence of a finite longitudinal response $\chi_{xx}$. In panel (e), although the $S_y$ component maintains its parity in $k_x$ across most of the BZ, this symmetry is strongly modified near the EPs. Specifically, the local redistribution of the spin density around these singularities tends to restore an even-like parity in $k_x$ locally; therefore, the integration over the BZ leads to a suppression of the $\chi_{xy}$ response. Finally, for the $S_z$ component in panel (f), the regions of spin polarization are drastically reduced, with the spin intensity confined primarily within the interior region defined by the EPs. This spatial confinement and subsequent cancelation across the zone cause the $\chi_{xz}$ component to vanish.

\subsubsection{Vertex correction and impact of disorder}

\begin{figure}
    \centering
    \includegraphics[width=.90\linewidth]{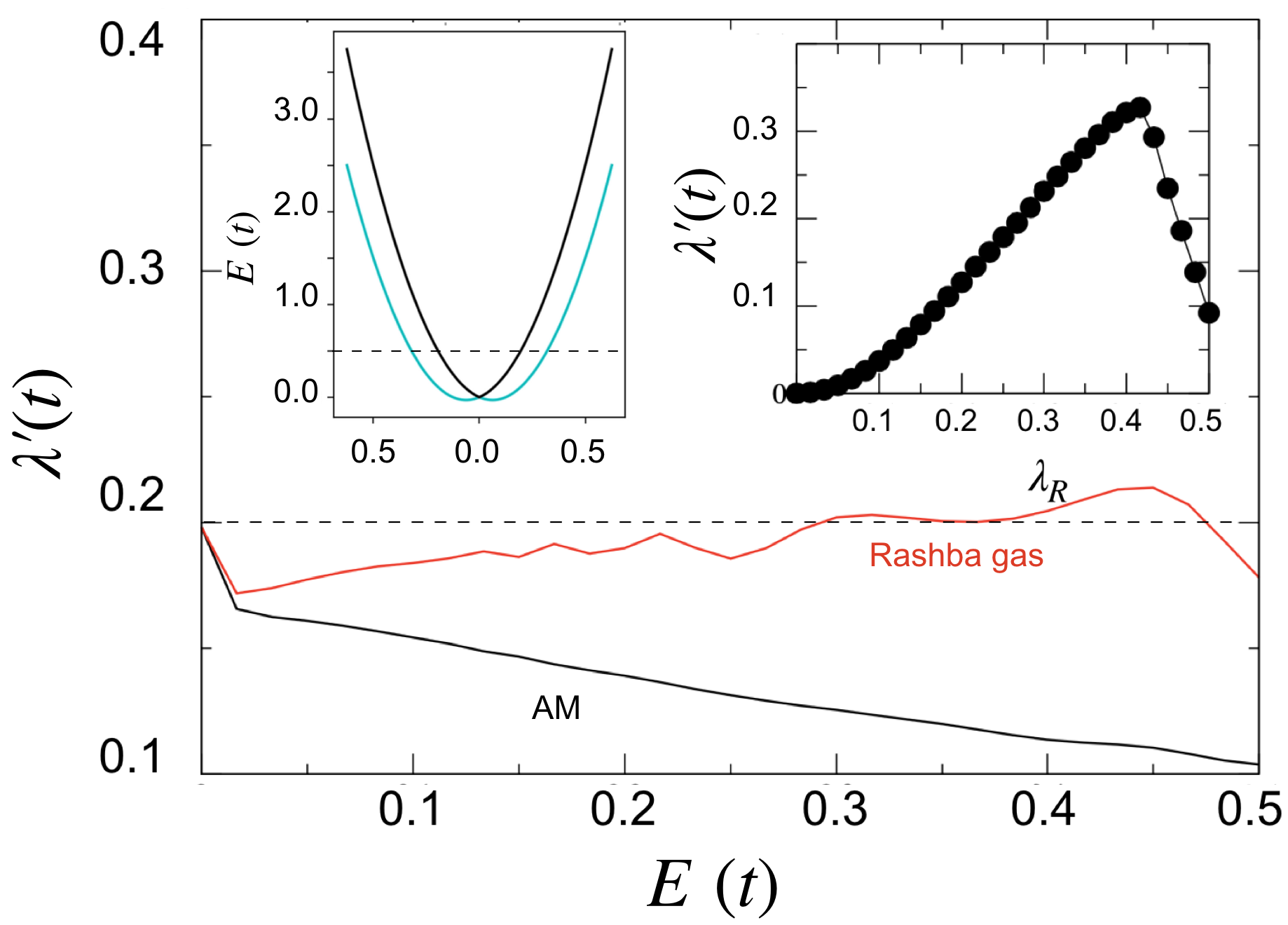}
    \caption{Correction to the parameter $\lambda_R$ denoted as $\lambda^{'}$ obtained from the vertex corrections when considering an impurity potential of 0.01t as a function of the Fermi level. Each of them was obtained by considering the full Altermagnet Hamiltonian having $\alpha=0.15t$ and $\lambda_R=0.2t$ (black) and pure Rashba Hamiltonian ($\alpha=0$) (red), where interesting features are displayed as the monotonic behavior of the black curve directly related to the existence of a Rashba effect in an altermagnet. In contrast, the red curve shows an almost constant behavior accounting for the exact cancellation of the velocity operator in the case of the Rashba case. The inset displays the band structure of the full model along the $k_x$ direction (left) and the modification of the correction $\lambda^{'}$ as the value of the Rashba parameter is modified on the right hand side.}
    \label{Fig:vertex}
\end{figure}

Since disorder and dirty are ubiquitous in materials, in this part, we comment briefly on their impact by considering the following standard expression $V(\mathbf{r})=V_0\sum_i\alpha(\mathbf{r}-\mathbf{r}_i)$ accounting for a non-magnetic potential distributed in random points $\mathbf{r}_i$, in this sense, we restrict our analysis to the correction of the spin operator $\hat{s}_y$, appearing in the velocity operator ($\partial \varepsilon_{n,\mathbf{k}}/\partial k_x$), by solving the Bethe-Salpeter equation in the ladder approximation \cite{PhysRevB.103.085401} 

\begin{equation}
    \tilde{s}_y=\hat{s}_y+n_iV_0\sum_kG^A(\mathbf{k},E)\tilde{s}_yG^R(\mathbf{k},E),
    \label{eq:vertex}
\end{equation}

\noindent being $n_i$ the density of impurities and $V_0$ the potential strength taken as a constant through all the sample. The low-energy dressed Hermitian GF evaluated at the energy $E$ corresponding to the long-wavelength limit ($|\mathbf{k}|\rightarrow0$) of the Hamiltonian ($H_{kp}(\mathbf{k})$) derived from Eq. \ref{hamiltonian_altermagnet} with $\Gamma = \gamma = 0$. For solving such a system, we assume a recursive procedure where the spin operators are written on the same basis of the Hamiltonian considered above, arriving at the following expression

\begin{equation}
    G^{R/A}(\mathbf{k},E_F) = \frac{1}{E_F-H_{kp}(\mathbf{k}) \pm i\eta},
\end{equation}

\noindent then by plugging it into Eq. \ref{eq:vertex}, we calculate the corrections to the coefficient $\lambda_R$, labeled as $\lambda'$, which is displayed in Fig. \ref{Fig:vertex} for an energy window surrounding band crossing. 

Unlike the $k$-independent velocity $\propto \sigma_y$ of the Ferromagnetic Rashba gas, our system exhibits an out-of-plane velocity component due to the helicity of the altermagnet with anisotropic spin-split bands. As shown in Fig. \ref{Fig:vertex}, this modifies the spin operator $\sigma_y$. The black curve (for AM coupling $0.15t$) shows a correction that depends on $\lambda_R$, reaching a maximum near $0.4t$ and then decreasing toward zero as $\lambda_R$ increases. This contrasts with the usual Rashba gas, where cancellation in $\partial H_R/\partial k_x \sim \sigma_y(\lambda_R-\lambda')=0$ yields the standard result \cite{PhysRevLett.97.046604,PhysRevB.70.041303}. On the other hand, by considering the effective $d$-wave AM Hamiltonian Eq. \ref{hamiltonian_altermagnet} with $\theta =0$, the velocity operator $\partial H_{AM}/\partial k_z$ includes a term $\propto \alpha k_y \sigma_z$,  preventing this cancellation, then leading to a qualitatively different behavior. This analysis has been reported already in the context of the anomalous and spin Hall effect in AMs \cite{Chen2025_ver}. In such a case, however, the correction diminishes away from the band crossing point (so called Dirac point), suggesting only a moderate or small effect on the full NH response as such corrections enter into Eq. \ref{extrinsic} modifying the velocity operator. 

\section{Conclusions}
\label{sec4}

We investigate the interplay between altermagnetism (unconventional magnetism) and NH dynamics in the context of the Edelstein effect. By considering an hybrid system comprising an AM(UM) coupled to a ferromagnetic lead, we demonstrate that the dissipative landscape, introduced via complex self-energies, acts as a symmetry-selective filter for spin-charge conversion. Our results reveal that while the transversal in-plane components of the spin susceptibility undergo significant damping due to the NH terms in both $d$-wave and $p$-wave symmetries, this dissipation facilitates a non-trivial redistribution of the spin accumulation. Specifically in the $d$-wave AM, we find the emergence of finite longitudinal and out-of-plane susceptibility components that are otherwise suppressed in the Hermitian limit. Moreover, these components arise for specific N\'eel vector orientations, in particular, the emergence of the diagonal component $\chi_{xx}$ for the $d_{xy}$-wave AM requires a finite altermagnetic parameter $\alpha$, which underscores the fundamental difference between this system and a conventional Rashba semiconductor. This phenomenon suggests that non-Hermiticity can be exploited to manipulate the spin-texture orientation in AMs, offering new knobs for realizing chiral spin-orbitronics features in dissipative quantum regimes. Our findings reveal that NH physics remarkably change the behavior of dynamical quantum properties in magnetic systems as also pointed out by \cite{hurst2022}. Specifically, we can propose the experimental realization of AM(UM)/FM lead junctions by considering an explicit coupling to an electron reservoir \cite{gordeeva2020}. In this framework, the damping of the physical signals arises from the non-conservative exchange with the environment, in the similar manner reported by Josephson junctions \cite{qi2025}. Given the recent experimental confirmation of the large altermagnetic spin-splitting in materials such as MnTe \cite{Krempasky2024} or LuFeO$_{3}$ which was reported to present a transverse susceptibility around $0.5\hbar \SI{}{\angstrom}/V$ without NH features \cite{chakraborty2025}, clearly, our model provides a realistic pathway for the experimental realization of spin-charge conversion devices where dissipation acts as a functional tuning parameter and this value is expected to decrease in an open environment. On the side of the $p$-wave candidate to prove our model, we can mention CeNiAsO, which reveals the value of $\chi_{xy} \approx 13\hbar \SI{}{\angstrom}/V$, which will decrease with the action of NH parameters.


\acknowledgments
 
The authors thank A. Manchon for helpful discussions and suggestions. This work was supported by the ANR 'ORION' project, grant ANR-20-CE30-0022-01 of the French Agence Nationale de la Recherche, ANR-22-CE30-0026 'DYNTOP' project, by a France 2030 government grant managed by the French National Research Agency PEPR SPIN ANR-22-EXSP0009 'SPINTHEORY' and  by the program “Excellence Initiative - research university” for AGH University of Krakow.
\bibliography{references}


\end{document}